# ON THE POSSIBILITY OF SEISMIC ROGUE WAVES IN VERY SOFT SOILS

Stéphane BRÛLÉ[1], Stefan ENOCH[1], Sébastien GUENNEAU[2]
[1] *Aix Marseille Univ, CNRS, Centrale Marseille, Institut Fresnel, Marseille, France*
[2] *UMI2004 Abraham de Moivre, CNRS-Imperial, London SW7 2AZ, United Kingdom*

**ABSTRACT** – Liquid wave studies have shown that under certain constructive interference conditions, an abnormal size wave could be generated at a specific point. This type of wave, described long ago in the literature, was named "Draupner Wave" in 1995. It was the first rogue wave, more than ten meters high, to be detected by a measuring instrument, occurring at the Draupner platform in the North Sea off the coast of Norway. For very soft and thick soils in continental sedimentary basin, with high water content, low shear wave velocity and wave bouncing effects on the edges of the basin, one can question the theoretical existence of such a type of freak waves that may have come unnoticed amongst surface seismic waves. This short communication presents recent observations and the conditions and limits of the analogy between gravity and seismic waves.

## 1. Introduction

In this short communication, we propose to open a complementary analysis of the complex wave propagation effects in the continental sedimentary basins and more precisely in very soft and thick deposits of soils.

Our proposal is motivated by recent research advances in the control of waves in structured media that make use of correspondences between various wave systems. Indeed, it is sometimes tempting to draw some parallels between liquid waves and seismic waves, since the former obey a simpler governing equation (typically scalar partial differential equation) that might be easier to handle analytically and numerically than a full elastodynamic model.

Advances in seismic full-wave modelling (Igel, 2017) help us to identify a large range of wave phenomena, expected and predictable, but it is not so obvious to claim in seismology and earthquake engineering. The main reason for this difficulty is dimensions of the phenomena involved, which are not necessarily observable on a human scale.

It is geotechnical data that leads us to reconsider the analogy of seismic surface waves in basins with gravity waves. For example, some soils may contain more than three times the weight in water than that in solid grain, making us think of an almost liquid material (Auvinet et al., 2017). A rare movie showing a plane parked on the tarmac at the Mexico City airport, during the earthquake of 19 September 2017 (magnitude $M_w$ 7.1), reinforced our conviction. We can indeed observe the horizontal and vertical cyclic displacement of the aircraft, which is analogous to the motion of a boat subjected to water waves (AFPS, 2018).
Another spectacular movie exists and has been recorded showing the horizontal metric displacement of the ground in the axis of a long street during the 2015 Gorkha earthquake (Gautam et al., 2016) in Nepal (magnitude $M_w$ 7.8, 25 April 2015). Part of the city rests on a former drained lake with soft sediments. In both cases, because of soft soils, the dominant wavelengths are visually perceptible (see web links in §7 References).

The plan of the short communication is as follows: First we give arguments in favor of an analogy between Rayleigh and gravity waves (§2) and remember the most common way in civil engineering to evaluate the ground response under seismic disturbance (§3). Then we recall the physical and mechanical properties of soft soils and the wave bouncing effects in sedimentary basin (§4). The differences and similitudes between terrestrial and gravity waves are presented in §5. We finally draw some concluding remarks and perspectives in §6.

## 2. Arguments for the analogy between seismic and gravity waves

Such analogies make possible useful qualitative predictions. For instance similar analogies drawn between electromagnetic waves propagating in media structured on the micrometer scale and surface seismic waves propagating in soils structured on a meter scale have led to an in-situ experiment (Brûlé et al., 2017a) demonstrating an unprecedented control of Rayleigh waves: some of the frequencies of a seismic signal generated by 17 ton mass drop near an array of boreholes have been focused on the other side of the array, according to the law of negative refraction in optics (Veselago, 1967).

We stress that the discovery of the focussing effect of seismic waves by a flat lens in (Brûlé et al., 2017a) was guided by observation of same phenomenon in optics (Gralak et al., 2000, Pendry, 2000), and since many other exciting effects have been observed in recent years in optical systems, it is likely that some of these will be achieved also for seismic waves (Brûlé et al., 2019).

Transferring the knowledge of waves in electromagnetic metamaterials to seismic waves in structured soils has led to the emergence of seismic metamaterials (Brûlé et al., 2017a). It is thus legitimate to ask whether some spectacular wave phenomena already observed for liquid waves, could be also observed for seismic waves.

In oceanography, rogue waves are more precisely defined as waves whose height is more than twice the significant wave height (SWH). The latter are defined as the mean of the highest one-third of waves in a wave record.

In Civil Engineering, a seismic wave locally of amplitude twice as large as the amplitude of surrounding soil vibrations does not necessarily produce a spectacular visual effect on the soil surface as in the case of the metric gravity waves. Indeed, the amplitude of the seismic displacement is of the order of a few millimetres to decimetres, but that can be enough to render inefficient the seismic base design of a building.

It is less complicated to simulate water wave propagation in wave pool (Figure 1) than to test a small-scale soil physical model under seismic disturbance. In fact, for water waves it could be a model testing at 1 g subject to conditions on the Reynolds number, while for the soil there are similitude requirements involving centrifugal acceleration (Brûlé et al., 2015).

If the soil is very soft with a very high water content, one can explore the analogies between terrestrial Rayleigh surface waves (Figure 1) and gravity waves.

## 3. Evaluation of ground response in civil engineering

One of the objectives in geotechnical earthquake engineering is the evaluation of ground response. The ground response analysis is involved in the process of determining the design response spectra for building by means of ground surface motions.

Ideally, a complete ground response analysis would model the rupture mechanism at the source location of an earthquake, the propagation of waves through the Earth to the top of bedrock beneath a particular site and would then determine how the ground motion is influenced by sedimentary deposits that lie above the seismic bedrock (Kramer, 1996).

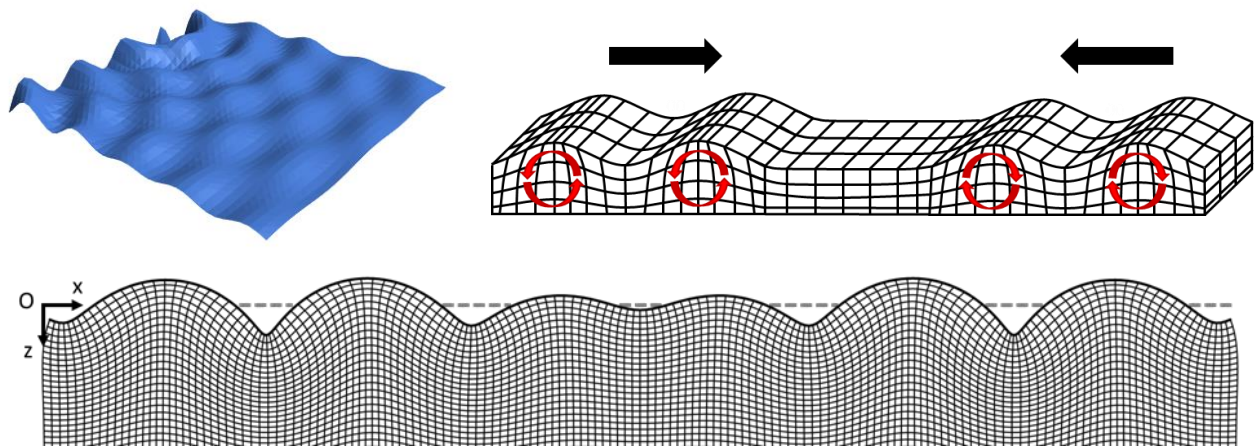

Figure 1. Principle of wave interferences. Top left: Interference between two linear waves generated by point sources simulated by sine functions (figure obtained by calculation sheet). Top right: Schematics of counter-directional propagation of two terrestrial Rayleigh surface wave fronts in homogeneous soil. Near the surface, the motion of particles is retrograde. Bottom: interference between two counter-propagating Rayleigh waves in a perfect elastic medium, with two plane fronts without relative incidence. The dotted gray line is the ground surface at rest.

In reality, the mechanism of fault rupture is complicated and the nature of energy transmission between the source and the site so uncertain that this approach is not practical for common engineering applications. In practice, empirical methods based on the characteristics of recorded earthquakes are used to develop predictive relationships used in conjunction with a seismic hazard analysis to predict bedrock motion characteristics at the studied site. Most of the time, the problem of ground response analysis becomes one of determining the response of the soil deposit to the motion of the bedrock immediately beneath it.

The influence of local soils conditions on the nature of earthquake damage has been accepted for many years and well described in few studies or post-seismic missions (AFPS, 2018). The ground motion amplification can be studied with a one-dimensional (1D) and two or three-dimensional (2/3D) wave propagation in a simplified soil model, commonly homogeneous or constituted of a stack of horizontally infinite layers for conditions of validity at the scale of a building or a district (Figure 2).

Very strong hypotheses are formulated such as the prior analysis of shear body waves rather than Rayleigh waves, the verticalization of the seismic ray up to the surface since the wave propagation velocity of shallower materials is generally lower than that of the materials beneath them, etc. (Brûlé et al., 2017a). For the future, we think that the Rayleigh waves should be taken into account.

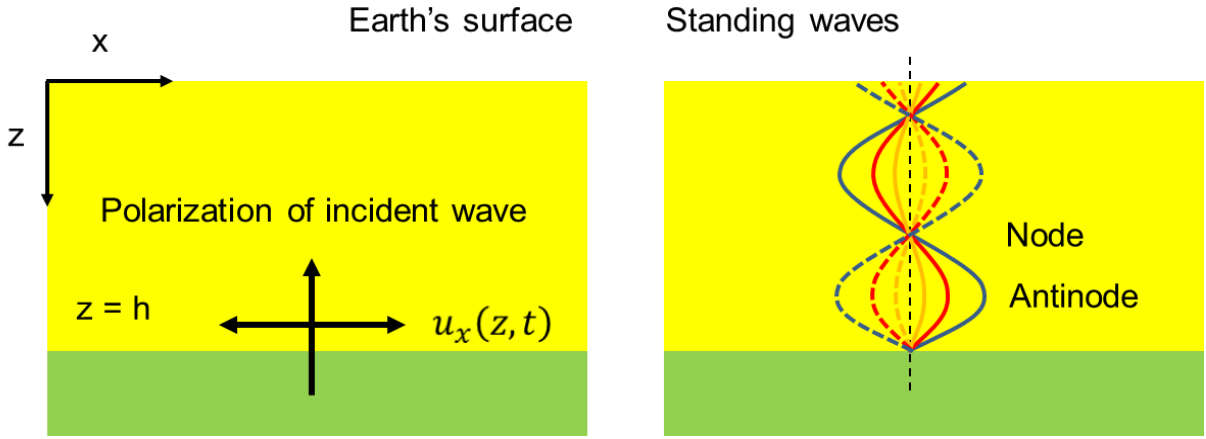

Figure 2. Schematics of wave propagation in a 1D soil-model. Left: The thickness of the upper homogeneous soil is $h$ (typically 10-100m) and the incidence of the plane-wave is vertical. The polarization is in the horizontal $(x, y)$ plane (shear S-wave). Right: Stationary wave pattern in the case of a phase interaction of the down-propagating reflected wave at the Earth's surface with the ascending one.

One-dimensional ground response analysis is based on the assumption that all boundaries are horizontal and that the response of a soil deposit is predominantly caused by SH-waves propagating vertically. In terms of displacement, that means $u_x(x, z, t) = u_x(z, t)$. Forward and reflected wavefields vibrate in phase to create constructive interferences. This allows to determine the period $T^0_{soil}$ (1) for the first fundamental mode, considering $h$, the thickness of the soil layer and $V_S$, the shear wave velocity:

$$T^0_{soil} = \frac{4h}{V_S} \qquad (1)$$

Procedures based on this assumption predicted a ground response in reasonable agreement with measured response in many cases. It is common to characterize the site effects by means of spectral ratios of recorded motions with respect to reference rock site (Cadet et al., 2010). These ratios, called empirical transfer functions, can be compared to theoretical curves obtained for a 1D-soil model (Figure 3).

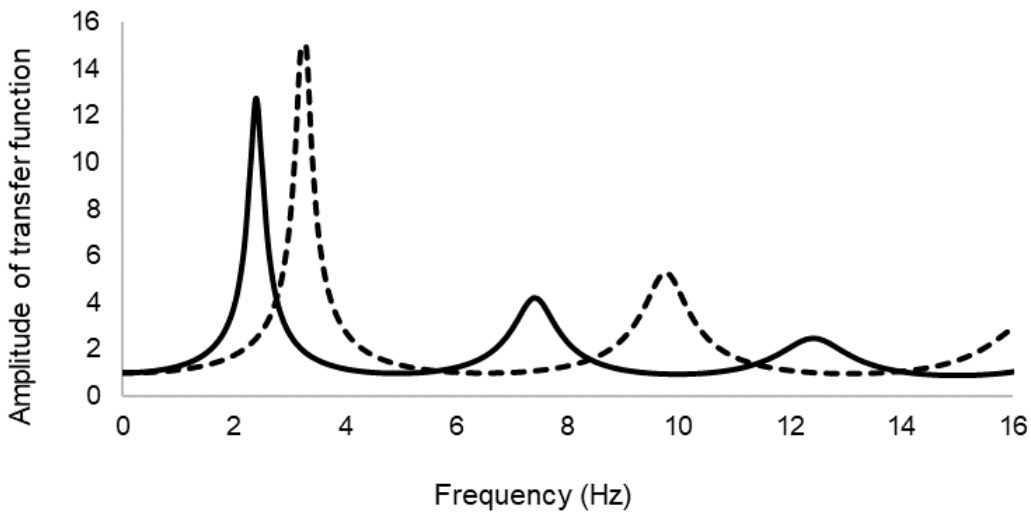

Figure 3. 1D soil-model. Transfer function for the theoretical displacement at the seismic substratum and the free surface. The thickness of soil deposits is 20 m. The shear wave velocity $V_s$ is 200 m.s$^{-1}$ (black solid line) and 260 m.s$^{-1}$ (black dotted line). The damping ratio is respectively 0.05 and 0.04 (inspired from Brûlé et al., 2019).

One way to try to explain the buildings' damage observed after an earthquake is to study the possibility of resonance between the seismic signal and the soil and the case of resonance between the soil and the building. Both types of resonance can occur simultaneously.

This is the hypothesis of "double-resonance" formulated for the city of Mexico during Puebla's Earthquake that occurred in September 19, 2017 (AFPS, 2018). However this analysis is often limited to comparing the frequency of the first fundamental mode and moreover, resonance phenomena do not necessarily explain all the destruction or damage of the structures at the soil surface. In general, it may be a lack of design, the non-respect of construction standards, the poor quality of the materials, an insufficient soil recognition to evaluate the right order of length of the site-amplification, etc.

Even if this widespread approach makes it possible to give a response to the objective of seismic design for buildings, it must not be forgotten that all wave phenomena can appear in terrestrial materials (Brûlé et al., 2014; Brûlé et al., 2017b), as in any other medium (Brillouin, 1946). But for that, it is necessary to sample the seismic signal by means of numerous sensors (Clayton et al., 2011) and sometimes their spacing should not exceed a few meters.

No one can dispute the fact that complex wave phenomena may exist and possibly be the cause of local amplification. To try to identify these phenomena, we can focus on those recently reproduced for scaled models for gravity waves and try to see how far the rheology of soft soils can reproduce these types of waves. Let us have a look at the description of soft soil.

**4. Soft soil and wave bouncing**

We choose to focus on the case study of Mexico City. The subsoil of the Mexico Valley is known worldwide for its lacustrine clays with high water content, sometimes up to 300%, (e.g. soils with three times more water weight than solid grain) and its seismic site effects, especially after the dramatic major earthquake of September 19th, 1985 (Auvinet et al., 2017). Typically, the geotechnical zonation of Mexico City consists of three areas: hills, transition and ancient lake zone. At the western part of the valley, the thickness of lacustrine deposits could reach 60 m. Shear wave velocity $V_S$ for the clays typically presents very low values (lower than 100 m/s).

Sedimentary basins surrounded by stiffer bedrock have a strong effect on earthquake ground motion because they trap and amplify seismic waves (Anderson et al., 1986). Basin effects can overwhelm the amplitude decay with distance from the earthquake source. The most dramatic example of strong seismic amplification distant from the source is the M 8.0 Michoacán, 1985, earthquake, which devastated Mexico City despite the rupture being located at a distance of about 300 km. This earthquake led the seismological community to pay closer attention to wave propagation effects in soft sediments (Bard et al., 1988; Aki, 1993).

On the other hand, seismic waves entering within a deep basin could be entrapped, bouncing from the edges and creating interference patterns that could enhance the ground motion at longer periods with an increase in the duration of the signal.

These waves rebound effects at the edges of the sedimentary basin and the likely consequences of interference, are at the origin of underpin this research paper.

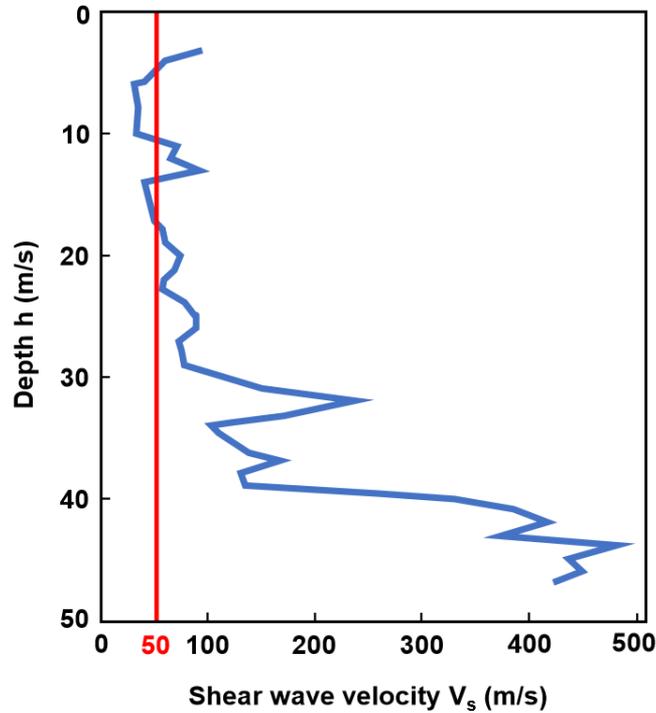

Figure 4. Velocity profile (blue) in the very soft soils of the Mexico Valley (from Roullé, 2004). Red solid line: typical 50 m/s mean value. Data from a borehole located at Colonia Roma (Plaza Morelia).

We relate the results from benchmarking numerical 3D methods of ground motion simulation in the valley of Grenoble in the French Alps (Chaljub et al., 2006; Tsuno et al., 2006), where large, broad band site effects occur. The benchmark consisted in computing the seismic response of the 'Y'-shaped Grenoble valley with local earthquakes (magnitude M≤3) for which recordings were available. Independently of the different numerical methods used (1D, 2D, 3D methods, prediction methods: Empirical Green's Functions (EGF), 2D Aki-Larner, 2D boundary elements, 2D and 3D Finite Differences, 3D Spectral Elements and 3D Discontinuous Galerkin Finite Elements), we stress that the energy mapping of trapped waves in the basin shows spots of intense energy as a consequence of wave interference phenomena.

## 5. Terrestrial Rayleigh waves and gravity waves

5.1 Gravity waves characteristics

The equations established by Euler in the 18th century describe quite well the waves. According to a simplified analysis of these equations by Lagrange, the speed of the waves does not depend on the wavelength (non-dispersive waves), contrary to a result of Newton published in its *Principiae Mathematicae* in 1687 (dispersive waves). A finer analysis has shown that the results of Newton and Lagrange are both valid, the first in deep water, the second in shallow water (Alvarez-Samaniego et al., 2008; Craik, 2004).

Let's examine the case of one-dimensional linear waves $u(x,t)$. The wave profile must satisfy the wave equation (1), which is a linear scalar partial differential equation (PDE):

$$\frac{\partial^2 u}{\partial t^2} - C_L^2 \frac{\partial^2 u}{\partial x^2} = 0 \qquad (2)$$

with $C_L$, the wave velocity.

Dispersions and nonlinear effects are an important aspect of wave physics. To take into account the nonlinearities, one can start by relying on the equations of Airy or Saint-Venant. By retaining only dominant nonlinearity, it is shown that the transport equation becomes (3):

$$\frac{\partial u}{\partial t} + C_L \left(1 + \frac{3u}{2H}\right) \cdot \frac{\partial u}{\partial x} = 0 \tag{3}$$

It is a nonlinear transport equation for which the velocity $C_L(1 + 3u/2H)$ depends on $u$. From there, several consequences are defined by Airy.

First, the velocity at the top of the wave is $\sqrt{gH}(1 + 3u_M/2H)$ but not $\sqrt{g(H + u_M)}$ with $u_M$ the maximum rise of the wave relative to the level of the water at rest. Secondly, there can be no wave propagating at constant velocity without being deformed because this would imply that velocity $C_L(1 + 3u/2H)$ takes the same value at any point of the wave (flat wave). Finally, as the top of the wave moves faster than the bottom, the waves break.

Airy's conclusions were a major theoretical and even practical advance, since they enabled him to explain the phenomenon of tidal bore, these tidal waves that trace the course of some rivers. The tidal wave is very "flat" (its period is 12 hours and its amplitude is a few meters at most), but the nonlinearities "stiffen" the wave when it goes up a river, creating this spectacular phenomenon.

To take into account the dispersion, i.e. waves of different wavelengths propagate at different phase velocities, we consider the transport equation of Korteweg-de Vries (4), which is a nonlinear and dispersive PDE:

$$\frac{\partial u}{\partial t} + C_L \left(1 + \frac{3u}{2H}\right) \cdot \frac{\partial u}{\partial x} + \frac{C_L H^2}{6} \frac{\partial^3 u}{\partial x^2} = 0 \tag{4}$$

The dispersion leads to vanishing waves' amplitude since their components separate while moving at different velocities. Conversely, nonlinearity tends to stiffen the waves and to make them break. The simultaneity of these two antagonistic phenomena neutralizes their respective effects and allows the wave to propagate without deforming itself (Craik, 2004).

5.2 Rayleigh waves characteristics

Surface waves are produced by body waves in media with a free surface and propagate parallel to the soil surface. For an epicenter remote from the zone of interest, the body waves emitted by the seismic source (shear and compression waves) are modelled as plane waves because of their large radius of curvature. The amplitude of surface wave decreases quickly with depth. For our purpose, we only consider an elastic, homogeneous half-space made of soils.

For plane waves, equations governing the displacement components of soil's particle $u_x$ and $u_z$ are described as follows (5 and 6), with $V_S$, $V_P$ and $V_R$, respectively the shear, pressure and Rayleigh waves velocities, $A$, the term of amplitude and nu, the Poisson's ratio, the wave number $k = 2\pi/\Lambda = \omega/V_R$ and $\omega$ the circular frequency:

$$u_x = \frac{i\omega}{V_R} A \left(e^{az} - \frac{2ab}{b^2 + \frac{\omega^2}{V_R^2}} e^{bz}\right) e^{i\frac{\omega}{V_R}(x - V_R t)} \tag{5}$$

$$u_z = aA \left(e^{az} - \frac{2\frac{\omega^2}{V_R^2}}{b^2 + \frac{\omega^2}{V_R^2}} e^{bz}\right) e^{i\frac{\omega}{V_R}(x - V_R t)} \tag{6}$$

Rayleigh waves are polarized in the vertical plane $(x, z)$. The imaginary term $i$ in the expression of the horizontal displacement induces a phase shift of $\pi/2$ with the vertical displacement. If we substitute in (4) and (5) the values of t during a complete cycle (0 to $T$, where $T = 2\pi/\omega$ is the period), we obtain for the particle's motion an ellipse with a vertical major axis and retrograde motion, opposite to that of wave propagation (Figure 1, right and Figure 5).

5.3 To sum up

The table below (Figure 5) shows the characteristics of the differential equations at play in gravity waves and seismic waves phenomena (inspired from Benzoni-Gavage, 2014). The notation $u$ represents an unknown (scalar or vector displacement fields or vector velocity field), $\sigma$ the Cauchy stress tensor (depending on $u$) and $p$ a pressure field. $f$ represents a forcing (the source). The other notations correspond to physical parameters. Symbol $\mu$ is the shear modulus or viscosity for elastic or fluidic media, respectively, and $\lambda$ is the compressional modulus, $\rho$ is the mass density, $h$ is related to the considered mass and the Planck constant, $k$ is related to the wavenumber (see also Figure 6) and $c$ is the wavespeed. Differential operators Laplacian $\nabla^2$, divergence $\nabla \cdot$ and gradient $\nabla$ act on space variable only.

| | Linear PDE | | Nonlinear PDE |
|---|---|---|---|
| **Type** | **Elliptic** | **Hyperbolic** | **Nonlinear and Dispersive** |
| **Type of problem** | Stationary problems | Evolution problems with finite velocity propagation | Hamiltonian evolution problems |
| **Linear Scalar PDE** | Helmholtz : $$\Delta u + k^2 u = 0$$ | Transport (gravity waves): First-order $$\partial_t u + a \cdot \nabla u = 0$$ Waves : $$\partial_{tt}^2 u - c^2 u = 0$$ | Korteweg-de Vries (gravity waves): $$\partial_t u + c\partial_x u + k\partial_{xxx}^3 u = 0$$ Boussinesq (gravity waves): $$\partial_{tt}^2 u - c^2 \partial_{xx}^2 u + k\partial_{xxxx}^4 u = 0$$ |
| **System of PDEs** | Stokes : Incompressible viscous fluid $$\begin{cases} -\mu \Delta \boldsymbol{u} + \nabla p = \boldsymbol{f} \\ \nabla \cdot \boldsymbol{u} = 0 \end{cases}$$ | Elastodynamics (seismic waves): $$\rho \partial_{tt}^2 \boldsymbol{u} - \nabla \cdot \sigma(\boldsymbol{u}) = f$$ For isotropic and homogeneous material : Navier-Cauchy $$\rho \partial_{tt}^2 \boldsymbol{u} = \mu \nabla^2 \boldsymbol{u} + (\mu + \lambda)\nabla(\nabla \cdot \boldsymbol{u}) + \boldsymbol{f}$$ | Schrödinger (gravity waves): $$ih\partial_t u = \Delta u$$ |

Figure 5. Synoptic table on PDEs at play in gravity waves and seismic waves phenomena (adapted from Benzoni, http://math.univ-lyon1.fr/~benzoni/EDP.pdf).

In order to better illustrate the striking similarities, and differences, between some of the types of waves governed by PDEs in Figure 5, we show in the diagram below (Figure 6) the main differences between surface seismic waves and gravity waves.

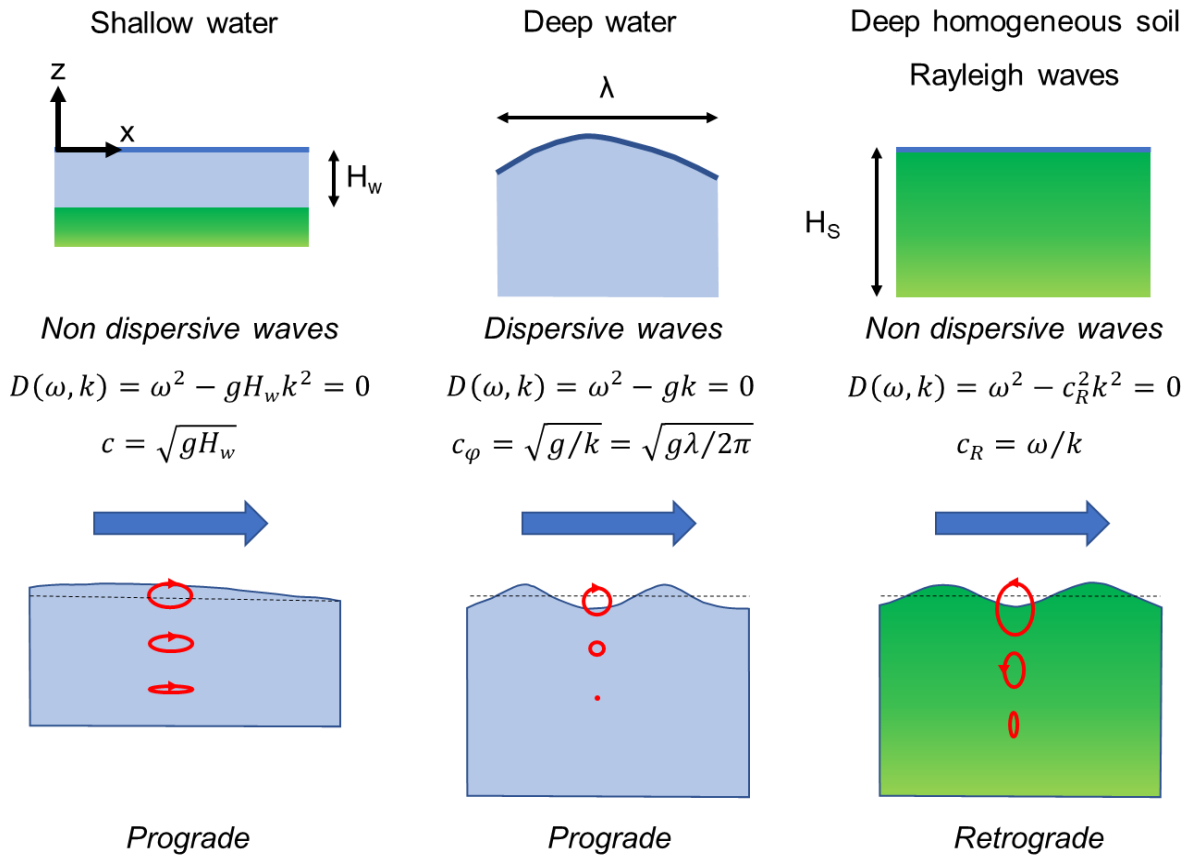

Figure 6. Comparison of dispersion relation and particle motion between gravity waves in deep and shallow water for linear conditions and Rayleigh surface waves propagating in homogeneous soil material.

It is apparent from this Figure 6 that there might be some soil configurations for which similar wave phenomena to those observed in shallow and deep water could occur. As already mentioned, the focussing effect of seismic waves by a flat lens in (Brûlé et al., 2017a) was guided by observation of same phenomenon in optics (Gralak et al., 2000, Pendry, 2000). These analogies were made possible using the elliptic Helmholtz equation in a context of time-harmonic, linear, wave phenomena (see first column in Figure 5). A striking example is that of Love waves propagating within a forest of trees atop of a soil substrate with a thin and soft guiding layer, which are governed by the exact same equation as electromagnetic surface waves, known as spoof plasmon polaritons, propagating in a structured metal (Maurel, 2018). Such analogies can be pursued in the time domain using hyperbolic acoustic and elastodynamic wave equations depending upon whether the seismic wave is out-of-plane or in-plane (see second column in Figure 5). Of course, a rigorous model of seismic waves should always be based on the fully vectorial Navier-Cauchy equations in the linear regime. However, the most devastating surface seismic waves operate in a non-linear regime, in which case the Korteweg-de Vries equations, which are routinely used in models of non-linear optical fibres, allow us to extend the analogy between seismic and water waves to terra incognita. One rare and extraordinary event we would like to investigate now is that of rogue waves. The question we ask ourselves is about a possible existence of such rare events in the context of surface seismic waves propagating in very soft soils.

5.2 Draupner wave

The 25.6 m high Draupner wave recorded in the North Sea on 1 January 1995 (Haver, 2004) and observed in a sea state with a significant wave height of 12 m, was one of the first confirmed field measurements of a freak wave. The physical mechanisms that give rise to freak waves such as the Draupner wave are still in discussion. There is obviously the Peregrine solution to the model of deep water waves using the nonlinear Schrödinger equation (Peregrine, 1983), with the first observation of a Peregrine solitary wave in a long water channel (Chahchoub et al., 2011). Such rogue waves have been also observed in nonlinear optical fibres (Kibler et al., 2010). However, rogue waves occur in the open sea and thus 1D models are not completely satisfactory for some research groupings. In actuality, there are alternative models and other experiments on the topical subject of freak waves. We focus here on recent experiments carried out in a circular wave tank, in which researchers attempt to recreate a freak wave (McAllister et al., 2019).

Of the possible mechanisms, the two general theories that prevail in the absence of specific environmental forcing are random dispersive focusing enhanced by weak bound-wave nonlinearity on the one hand and the modulational instability of waves trains in deep water on the other (McAllister et al., 2019). The latter applies mainly to sea states that are narrow banded in both frequency and direction and sufficiently deep (typically, $kH_w \gg 1.36$, where $k$ is the wavenumber and $H_w$ the water depth).

5.3 Interference with two Morlet wavelets

We use a calculation sheet to generate two Morlet wavelets with wavefronts making an angle of 120 degrees and we observe the occurrence of rogue waves in Figure 7. The assumptions made in this simple simulation are linearity and no dispersion. We argue that these conditions can simulate the case of Rayleigh surface waves, which can be encountered in soft soils with high water concentration such as in the Mexico city sedimentary basin. Along the points of maximum amplitude, damage alignments could possibly be identified.

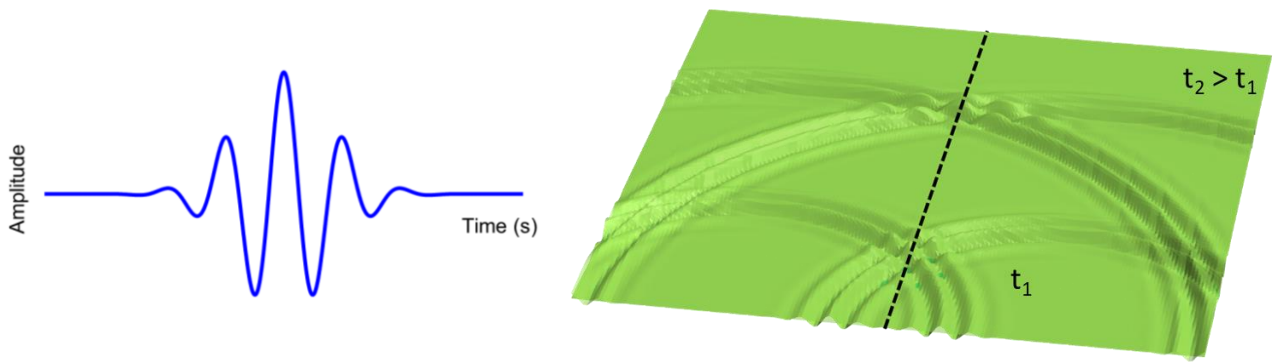

Figure 7. Morlet wavelet. Left: typical shape. Right: Interference between two linear Morlet wavelets at two different times ($t_2$ and $t_1$). The angle made between the normal lines to the two wavefronts is 120°. The dotted black line symbolically represents the maximum of the amplitude over times.

**6. Discussion and perspectives**

In this short communication, we propose a concept of rogue waves for surface seismic waves in sedimentary basins with high-concentration of water such as in the sedimentary basin of Mexico City. We do not model such waves as solutions of nonlinear Schrödinger equations but rather use the model of two Morlet wavelets borrowed from a recent study (MacAllister et al. 2019). We realise that our model can be improved but we hope that our proposal can foster theoretical and experimental efforts in rogue-like surface seismic waves.

More measurements on denser meshes of sensors will provide us with unexpected information. Progress can be focused on identifying a specific energy distribution inside structured soils (buried structures, foundations, etc.) or on detecting interferences in sedimentary basins or many over phenomena. In the same spirit of a better acuity of the observations, we can quote the following case, recent, showing the wealth of the seismological information.

Recent seismic events had shown the complex relationship between seismic data acquisition process and interpretation of the results. For example, on november 11, 2018 around 9:30 am (UTC), seismographs from around the world, up to 11 000 kilometers from Mayotte (French island, part of Comoros Archipelago in Indian Ocean) picked up seismic waves that were predominantly in a frequency range far below that of normal earthquakes. The classic pressure (P) and shear (S) waves were very weak in magnitude compared to surface waves. The signal repeated about every 17 seconds (period of 17 s or 0.06 Hz) and was almost monochromatic.

Normally, this type of long-range waves is generated by earthquakes of great magnitude (greater than M 5 and shallow depths <30 km). The source of these waves appears to be a swarm zone. The mechanism that triggered the wave is so far unexplained. The hypotheses advanced by researchers, such as magma chamber resonance, have yet to be confirmed by additional data. However, the phenomenon observed seems to be a further indication of a volcanic component in the ongoing earthquake swarm (BRGM website, 2018).

What we suggest with this short communication is that the sedimentary basins that are real traps to waves, meet the conditions necessary to observe phenomena similar to those visible for gravity waves.

Based on recent observations, this simple analogy enables us to conjecture existence of nodal points consecutive to constructive interferences and to progress in the explanation of the visible disorders on surface constructions. The next step may be to take into account, for the design, an even more precise definition of maximum soil displacements during earthquake, not based solely on a 1D model for bulk shear waves (horizontal amplification), but taking into account surface waves and any constructive interferences.